\documentclass[english,aps,preprint,superscriptaddress]{revtex4}
\usepackage[]{fontenc}
\usepackage[latin9]{inputenc}
\usepackage{amsmath}
\usepackage{esint}
\usepackage{epsfig}
\usepackage{graphicx}
\usepackage{bm}
\usepackage{rotating}
\makeatletter
\@ifundefined{textcolor}{}
{%
 \definecolor{BLACK}{gray}{0}
 \definecolor{WHITE}{gray}{1}
 \definecolor{RED}{rgb}{1,0,0}
 \definecolor{GREEN}{rgb}{0,1,0}
 \definecolor{BLUE}{rgb}{0,0,1}
 \definecolor{CYAN}{cmyk}{1,0,0,0}
 \definecolor{MAGENTA}{cmyk}{0,1,0,0}
 \definecolor{YELLOW}{cmyk}{0,0,1,0}
 }

\makeatletter

\newcommand{\be}{\begin{equation}}\newcommand{\ee}{\end{equation}}\newcommand{\ba}{\begin{align}}\newcommand{\ea}{\end{align}}\def\bea{\begin{eqnarray}}\def\eea{\end{eqnarray}}

\usepackage{slashed}

\makeatother

\usepackage{babel}

\begin{document}

\title{Neutrino mass textures with one vanishing minor and two equal cofactors}

\author{Weijian Wang}

\affiliation{Department of Physics, North China Electric Power
University, Baoding 071003, P. R. China} \affiliation{Zhejiang
Institute of Modern Physics, Zhejiang University, Hangzhou 310027,
P.R. China}

\email{wjnwang96@gmail.com}

\begin{abstract}
In this paper, we carry out a numerical and systematic analysis of
the neutrino mass textures, which contain one vanishing minor and an
equality between two cofactors. Among 60 logically possible
textures, only eight of them are excluded for both normal and
inverted hierarchy by the current experimental data at 3$\sigma$
level. We also demonstrate that the future long-baseline neutrino
oscillation experiments, especially for the measurement of
$\theta_{23}$ mixing angle, will play the important role in the
model selection. The phenomenological implications from neutrinoless
double beta decay and the cosmology observation are also examined. A
discussion on the flavor symmetry realization of the textures is
also given.
\end{abstract}
\maketitle

\section{Introduction} By this time, neutrino oscillation experiments
have provided us with convincing evidences for massive neutrinos and
leptonic flavor mixing. The bi-large flavor structure with two large
mixing angle
($\theta_{12}\approx34^{\circ}$,$\theta_{23}\approx38^{\circ}$) and
the mass-squared differences ($\delta m^{2}\equiv
m_{2}^{2}-m_{1}^{2}$ and $\Delta m^{2}\equiv \mid
m_{3}^{2}-(m_{1}^{2}+m_{2}^{2})/2\mid$) have been revealed by solar,
atmosphere and accelerator oscillation experiments with high degree
of accuracy\cite{neu1}. Recently, the third mixing angle
$\theta_{13}$ has been precisely measured by Daya Bay\cite{neu2} and
RENO\cite{neu3} experiments. The relative large value of
$\theta_{13}$($\approx 9^{\circ}$)opens the door for us to explore
the leptonic CP violation and determine the neutrino mass hierarchy
in the future long-baseline oscillation experiments. Although the
absolute mass scale and the type of neutrino (Dirac type or Majorana
type) is still unknown, they are constrained by the cosmology
observation\cite{WMAP, Planck} and neutrinoless double-beta decay
experiments (for a review, see \cite{NDD}). From the point of model
building, it is important to find the appropriate textures of
leptonic mass matrices consistent with the current experimental
data. Many ideas have been made to explore the phenomenologically
acceptable textures, such as texture zeros\cite{zero}, hybrid
textures\cite{hybrid, hybrid2}, zero trace\cite{sum}, zero
determinant\cite{det}, vanishing minors\cite{minor}, two traceless
submatrices\cite{tra} Recently the patterns with two equalities
between elements and cofactors\cite{co} have been investigated in
$3\times3$ neutrino mass textures. Furthermore, it is pointed that
some of the patterns, not all of them though, can be naturally
realized by introducing proper flavor symmetry in the framework of
seesaw mechanism.

In this paper, we systematically study the pattern of $M_{\nu}$ with
one vanishing minor and an equality between cofactors. In the type-I
seesaw model\cite{seesaw}, the effective neutrino mass matrix is
given by $M_{\nu}\approx M_{D}M_{R}^{-1}M_{D}^{T}$ where $M_{D}$ and
$M_{R}$ are the Dirac neutrino mass and right-handed Majorana
neutrino mass matrix respectively. In the basis where $M_{D}$ is
diagonal, the vanishing minor in $M_{\nu}$ is equivalent to the zero
textures of $M_{R}$\cite{Ma}. If $M_{D}$ is assumed to be
proportional to unit matrix, as having been made in Ref\cite{co}, an
equality between two cofactors can be seen as two equal elements in
$M_{R}$. However, as we will shown in the following section, in some
cases the unit $M_{D}$ condition can be loosen to the diagonal
matrix with two unit elements. Anyway, we are effectively studying
the cases with one zero element and two equal elements in
$M^{-1}_{\nu}$. In the following, we assume the neutrinos to be
majorana particles where the mass matrix $M_{\nu}$ is a $3\times3$
complex symmetric matrix. If one of elements of $M_{\nu}^{-1}$ is
zero and two of the others are equal, we have $C^{1}_{6}\cdot
C^{2}_{5}=60$ logically possible patterns.

The plan of the paper is organized as follows: in Sec. II, we show
standard formation for the three-flavor neutrino mixing and its link
to the experimental results. The numerical results are given in Sec.
III. We summary our results in Sec. IV, where a discussion on flavor
symmetry realization is also given.

\section{Formalism}
\subsection{Classification of The 60 Patterns }
 As having mentioned, we have 60 logically possible textures for the
mass matrix with one vanishing minor and two equal cofactor. Thus,
it is helpful for us to classify and label these patterns. According
to the zeros in the $M_{\nu}^{-1}$, we classified the textures into
six subgroups A-F as following
\begin{eqnarray}
&A1: C_{11}=0, C_{12}=C_{13};\quad A2:C_{11}=0, C_{12}=C_{22};\quad
A3:C_{11}=0, C_{13}=C_{23};\nonumber\\
&A4:C_{11}=0, C_{22}=C_{23};\quad A5:C_{11}=0, C_{22}=C_{33};\quad
A6:C_{11}=0, C_{23}=C_{33};\nonumber\\
&A7:C_{11}=0, C_{12}=C_{23};\quad A8:C_{11}=0, C_{13}=C_{33};\quad
A9:C_{11}=0, C_{13}=C_{22};\nonumber\\
&A10:C_{11}=0, C_{12}=C_{33};\nonumber\\
&B1: C_{12}=0, C_{11}=C_{13};\quad B2:C_{12}=0, C_{13}=C_{22};\quad
B3:C_{12}=0, C_{13}=C_{23};\nonumber\\
&B4:C_{12}=0, C_{13}=C_{33};\quad B5:C_{12}=0, C_{11}=C_{22};\quad
B6:C_{12}=0, C_{11}=C_{23};\nonumber\\
&B7:C_{12}=0, C_{11}=C_{33};\quad B8:C_{12}=0, C_{22}=C_{23};\quad
B9:C_{12}=0, C_{22}=C_{33};\nonumber\\
&B10:C_{12}=0, C_{23}=C_{33};\nonumber\\
&C1: C_{13}=0, C_{11}=C_{12};\quad C2:C_{13}=0, C_{11}=C_{22};\quad
C3:C_{13}=0, C_{11}=C_{23};\nonumber\\
&C4:C_{13}=0, C_{11}=C_{33};\quad C5:C_{13}=0, C_{12}=C_{22};\quad
C6:C_{13}=0, C_{12}=C_{23};\nonumber\\
&C7:C_{13}=0, C_{12}=C_{33};\quad C8:C_{13}=0, C_{22}=C_{23};\quad
C9:C_{13}=0, C_{22}=C_{33};\nonumber\\
&C10:C_{13}=0, C_{23}=C_{33};\nonumber\\
&D1: C_{22}=0, C_{11}=C_{12};\quad D2:C_{22}=0, C_{11}=C_{13};\quad
D3:C_{22}=0, C_{11}=C_{23};\nonumber\\
&D4:C_{22}=0, C_{11}=C_{33};\quad D5:C_{22}=0, C_{12}=C_{13};\quad
D6:C_{22}=0, C_{12}=C_{23};\nonumber\\
&D7:C_{22}=0, C_{12}=C_{33};\quad D8:C_{22}=0, C_{13}=C_{23};\quad
D9:C_{22}=0, C_{13}=C_{33};\nonumber\\
&D10:C_{22}=0, C_{23}=C_{33};\nonumber\\
&E1: C_{23}=0, C_{11}=C_{12};\quad E2:C_{23}=0, C_{11}=C_{13};\quad
E3:C_{23}=0, C_{11}=C_{22};\nonumber\\
&E4:C_{23}=0, C_{11}=C_{33};\quad E5:C_{23}=0, C_{12}=C_{13};\quad
E6:C_{23}=0, C_{12}=C_{22};\nonumber\\
&E7:C_{23}=0, C_{12}=C_{33};\quad E8:C_{23}=0, C_{13}=C_{22};\quad
E9:C_{23}=0, C_{13}=C_{33};\nonumber\\
&E10:C_{23}=0, C_{22}=C_{33};\nonumber\\
&F1: C_{33}=0, C_{11}=C_{12};\quad F2:C_{33}=0, C_{11}=C_{13};\quad
F3:C_{33}=0, C_{11}=C_{22};\nonumber\\
&F4:C_{33}=0, C_{11}=C_{23};\quad F5:C_{33}=0, C_{12}=C_{13};\quad
F6:C_{33}=0, C_{12}=C_{22};\nonumber\\
&F7:C_{33}=0, C_{12}=C_{23};\quad F8:C_{33}=0, C_{13}=C_{22};\quad
F9:C_{33}=0, C_{13}=C_{23};\nonumber\\
&F10:C_{33}=0, C_{22}=C_{23}; \label{hybridin}\end{eqnarray} where
the $C_{ij}$ denotes the cofactor of element $(i,j)$ in $3\times3$
neutrino mass matrix. For example, for the A1 texture with the
condition $C_{11}=0$, $C_{12}=C_{13}$, we have
\begin{equation}
M_{22}M_{33}-M_{23}M_{32}=0
\end{equation}
and
\begin{equation}
(-1)\cdot (M_{21}M_{33}-M_{23}M_{31})-(M_{21}M_{32}-M_{22}M_{31})=0
\end{equation}
On the other hand, we can also classify the textures in the
formation of $M_{\nu}^{-1}$, with one zero and two equal elements.
Then the A1 pattern, for example, can denote as
\begin{equation}
\left(\begin{array}{ccc}
0&\bigtriangleup&\bigtriangleup\\
\bigtriangleup&\times&\times\\
\bigtriangleup&\times&\times
\end{array}\right)
\label{hybrid}\end{equation} where the "$\bigtriangleup$" stand for
the nonzero and equal elements, while the "$\times$" stand for
arbitrary elements. The type of the matrix like \eqref{hybrid} has
appeared in the so-called hybrid neutrino mass texture\cite{hybrid}.
However, the situation here is totally different where the hybrid
feature shows not in $M_{\nu}$ but in $M_{\nu}^{-1}$; thus we can
call the texture given in \eqref{hybridin} the hybrid $M_{\nu}^{-1}$
textures.

\subsection{Standard Notation and Important Relations}
 In this section, we construct the neutrino mass texture
in terms of three neutrino mass eigenvalues ($m_{1}, m_{2}, m_{3}$),
three mixing angles($\theta_{12}, \theta_{23}, \theta_{13}$) and
three CP-violating phase($\delta, \alpha, \beta$). In the basis
where the charged mass matrix is diagonal, the neutrino mass texture
$M_{\nu}$ under flavor basis is given by\cite{Fog}
\begin{equation}
M_{\nu}=P_{l}VM_{diag}V^{T}P_{l} \label{MM}\end{equation} where
$M_{diag}$ is the diagonal matrix of neutrino mass eigenvalues
$M_{diag}=$diag$(m_{1}, m_{2}, m_{3})$ and $P_{l}$ denotes the
phases which are unobservable and depend on phase convention
\begin{equation}
P_{l}=\left(\begin{array}{ccc}
e^{i\phi_{e}}&0&0\\
0&e^{i\phi_{\mu}}&0\\
0&0&e^{i\phi_{\tau}}
\end{array}\right)
\end{equation}

The Pontecorvo-Maki-Nakagawa-Sakata matrix\cite{PMNS} $V$ can be
parameterized as $V=UP_{\nu}$ with
\begin{equation}
U=\left(\begin{array}{ccc}
c_{12}c_{13}&c_{13}s_{12}&s_{13}e^{-i\delta}\\
-s_{12}c_{23}-c_{12}s_{13}s_{23}e^{i\delta}&c_{12}c_{23}-s_{12}s_{13}s_{23}e^{i\delta}&c_{13}s_{23}\\
s_{23}s_{12}-c_{12}c_{23}s_{13}e^{i\delta}&-c_{12}s_{23}-c_{23}s_{12}s_{13}e^{i\delta}&c_{13}c_{23}
\end{array}\right)
\label{3}\end{equation}
and
\begin{equation}
 P_{\nu}=\left(\begin{array}{ccc}
1&0&0\\
0&e^{i\alpha}&0\\
0&0&e^{i(\beta+\delta)}
\end{array}\right)
\label{4}\end{equation}

Here the abbreviation $s_{ij}=\sin\theta_{ij}$ and
$c_{ij}=\cos\theta_{ij}$ is used. The $\alpha$ and $\beta$ in
$P_{\nu}$ denote two Majorana CP-violating phases and $\delta$
denotes the Dirac CP-violating phase. In neutrino oscillation
experiments, CP violation effect is usually reflected by the
Jarlskog rephasing invariant quantity\cite{Jas} defined as
\begin{equation}
J\equiv
Im(U_{e1}U_{\mu2}U^{*}_{e2}U^{*}_{\mu2})=s_{12}s_{23}s_{13}c_{12}c_{23}c_{13}^{2}\sin\delta
\end{equation}

Using \eqref{MM}, any element $M_{\nu(ab)}$in the neutrino mass
matrix can be expressed as
\begin{equation}
M_{\nu(ab)}=e^{i(\phi_{a}+\phi_{b})}\sum_{i=1}^{3}V_{ai}V_{bi}m_{i}
\label{el}\end{equation} The cofactors of $M_{\nu}$ are
\begin{eqnarray}
&(-1)^{m+n}(M_{\nu(ab)}M_{\nu(cd)}-M_{\nu(ef)}M_{\nu(gh)})\nonumber\\
&=(-1)^{m+n}\sum_{i,j=1}^{3}\big(e^{i(\phi_{a}+\phi_{b}+\phi_{c}+\phi_{d})}V_{ai}V_{bi}V_{cj}V_{dj}-e^{i(\phi_{e}+\phi_{h}+\phi_{g}+\phi_{h})}
V_{ei}V_{fi}V_{gj}V_{hj}\big)m_{i}m_{j} \label{cof}\end{eqnarray}
where $m$ and $n$ refer to the cofactor $C_{mn}$. It is observed
that for any cofactor there is an inherent property
\begin{equation}
\phi_{a}+\phi_{b}+\phi_{c}+\phi_{d}=\phi_{e}+\phi_{h}+\phi_{g}+\phi_{h}
\label{ins}\end{equation} Thus we can extract this total phase
factor from the bracket in Eq.  \eqref{cof}.

The textures we concerned have the property
\begin{equation}
M_{\nu(pq)}M_{\nu(rs)}-M_{\nu(tu)}M_{\nu(vw)}=0
\end{equation}
corresponding to the vanishing minor condition and
\begin{equation}
(-1)^{m+n}(M_{\nu(ab)}M_{\nu(cd)}-M_{\nu(ef)}M_{\nu(gh)})-(-1)^{m^{\prime}+n^{\prime}}(M_{\nu(a^{\prime}b^{\prime})}
M_{\nu(c^{\prime}d^{\prime})}-M_{\nu(e^{\prime}f^{\prime})}M_{\nu(g^{\prime}h^{\prime})})=0
\end{equation}
corresponding to the equivalent cofactor condition. Using
\eqref{el}, \eqref{cof} and the intrinsic property \eqref{ins}, we
obtain two equations for vanishing minor condition and equivalent
cofactor condition, i.e
\begin{equation}
m_{1}m_{2}K_{3}e^{2i\alpha}+m_{2}m_{3}K_{1}e^{2i(\alpha+\beta+\delta)}+m_{3}m_{1}K_{2}e^{2i(\beta+\delta)}=0
\label{c1}\end{equation}
\begin{equation}
m_{1}m_{2}L_{3}e^{2i\alpha}+m_{2}m_{3}L_{1}e^{2i(\alpha+\beta+\delta)}+m_{3}m_{1}L_{2}e^{2i(\beta+\delta)}=0
\label{c2}\end{equation} where
\begin{equation}
K_{i}=(U_{pj}U_{qj}U_{rk}U_{sk}-U_{tj}U_{uj}U_{vk}U_{wk})+(j\leftrightarrow
k)\end{equation}
\begin{eqnarray}
&L_{i}=(-1)^{m+n}Q(U_{aj}U_{bj}U_{ck}U_{dk}-U_{ej}U_{fj}U_{gk}U_{hk})\nonumber\\
&-(-1)^{m^{\prime}+n^{\prime}}
(U_{a^{\prime}j}U_{b^{\prime}j}U_{c^{\prime}k}U_{d^{\prime}k}-U_{e^{\prime}j}U_{f^{\prime}j}U_{g^{\prime}k}U_{h^{\prime}k})+(j\leftrightarrow
k)
\end{eqnarray}
with $(i,j,k)$ a cyclic permutation of (1,2,3) and $Q$ defined as
\begin{equation}
Q\equiv
e^{i\Phi}=e^{i(\phi_{a}+\phi_{b}+\phi_{c}+\phi_{d}-\phi_{a^{\prime}}-\phi_{b^{\prime}}-\phi_{c^{\prime}}-\phi_{d^{\prime}})}
\end{equation}
After solving Eq.\eqref{c1} and \eqref{c2}, we arrive at
\begin{equation}
\frac{m_{1}}{m_{2}}e^{-2i\alpha}=\frac{K_{3}L_{1}-K_{1}L_{3}}{K_{2}L_{3}-K_{3}L_{2}}
\label{r1}\end{equation}
\begin{equation}
\frac{m_{1}}{m_{3}}e^{-2i\beta}=\frac{K_{2}L_{1}-K_{1}L_{2}}{K_{3}L_{2}-K_{2}L_{3}}e^{2i\delta}
\label{r2}\end{equation}

With the help of Eq.\eqref{r1} and \eqref{r2}, we obtain the
magnitudes of mass radios
\begin{equation}
\rho=\Big|\frac{m_{1}}{m_{3}}e^{-2i\beta}\Big|
\label{t1}\end{equation}
\begin{equation}
\sigma=\Big|\frac{m_{1}}{m_{2}}e^{-2i\alpha}\Big|
\label{12}\end{equation} as well as the two Majorana CP-violating
phases
\begin{equation}
\alpha=-\frac{1}{2}arg\Big(\frac{K_{3}L_{1}-K_{1}L_{3}}{K_{2}L_{3}-K_{3}L_{2}}\Big)
\label{21}\end{equation}
\begin{equation}
\beta=-\frac{1}{2}arg\Big(\frac{K_{2}L_{1}-K_{1}L_{2}}{K_{3}L_{3}-K_{2}L_{3}}e^{2i\delta}\Big)
\label{22}\end{equation} The results of Eq. \eqref{t1},\eqref{12},
\eqref{21} and \eqref{22} imply that the two mass ratio ($\rho$ and
$\sigma$) and two Majorana CP-violating phases ($\alpha$ and
$\beta$) are fully determined in terms of three mixing angle and
Dirac CP-violating phase ($\theta_{12}, \theta_{23}, \theta_{13}$
and $\delta$). The neutrino mass ratios $\rho$ and $\sigma$ are
related to the ratios of two neutrino mass-squared ratios obtained
from the solar and atmosphere oscillation experiments as
\begin{equation}
R_{\nu}\equiv\frac{\delta m^{2}}{|\Delta
m^{2}|}=\frac{2\rho^{2}(1-\sigma^{2})}{|2\sigma^{2}-\rho^{2}-\rho^{2}\sigma^{2}|}
\label{rv}\end{equation} and to the three neutrino mass as
\begin{equation}
 m_{2}=\sqrt{\frac{\delta m^{2}}{1-\sigma^{2}}}\quad\quad
 m_{1}=\sigma m_{2}\quad\quad m_{3}=\frac{m_{1}}{\rho}
\label{abm}\end{equation} For normal neutrino mass hierarchy(NH),
the latest global-fit neutrino oscillation experimental data, at the
3$\sigma$ confidential level, is list as follows\cite{data}
\begin{eqnarray}
30.6^{\circ}\leq\theta_{12}\leq36.8^{\circ},
&35.1^{\circ}\leq\theta_{23}\leq53.0^{\circ},&
7.5^{\circ}\leq\theta_{13}\leq10.2^{\circ}
\end{eqnarray}
and
\begin{eqnarray}
6.99\times10^{-5}\textup{eV}^{2}\leq\delta
m^{2}\leq8.18\times10^{-5}\textup{eV}^{2}\nonumber\\
2.19\times10^{-3}\textup{eV}^{2}\leq\Delta
m^{2}\leq2.62\times10^{-3}\textup{eV}^{2} \label{ms}\end{eqnarray}
For the inverted neutrino mass hierarchy(NH), the differences
compared with the NH are so slight that we shall use the same values
given above. It is noted that the global analysis tends to give a
$\theta_{23}$ less than $45^{\circ}$ which is
\begin{equation}
36.2^{\circ}\leq\theta_{23}\leq42.0^{\circ}
\label{2sigma}\end{equation} at $2\sigma$ level and
\begin{equation}
37.2^{\circ}\leq\theta_{23}\leq40.0^{\circ}
\end{equation}
at $1\sigma$ level.

As has been pointed out by many papers\cite{mut}, the $\mu-\tau$
permutation symmetry between $2-3$ rows and $2-3$ columns of
$M_{\nu}$ can one patterns to another, i.e
\begin{equation}
\widetilde{M}_{\nu}=P_{23}M_{\nu}P_{23}
\end{equation}
where
\begin{equation}
P_{23}=\left(\begin{array}{ccc}
1&0&0\\
0&0&1\\
0&1&0
\end{array}\right)
\end{equation}
It is straight to prove that the $\mu-\tau$ permutation symmetry
leads to the following relation of mixing parameters between
$M_{\nu}$ and $\widetilde{M}_{\nu}$:
\begin{equation}
\widetilde{\theta}_{12}=\theta_{12},\quad
\widetilde{\theta}_{13}=\theta_{13},\quad
\widetilde{\theta}_{23}=\frac{\pi}{2}-\theta_{23},\quad
\widetilde{\delta}=\pi-\delta
\end{equation}
and $\widetilde{M}_{\nu}$ and $M_{\nu}$ have the same mass
eigenvalues. Here we list all the relations of textures related by
$\mu-\tau$ symmetry as
\begin{eqnarray}
&A1\leftrightarrow A1, A2\leftrightarrow A8, A3\leftrightarrow A7,
A4\leftrightarrow A6, A5\leftrightarrow A5, A9\leftrightarrow A10,
B1\leftrightarrow C1\nonumber\\
&B2\leftrightarrow C7, B3\leftrightarrow C6, B4\leftrightarrow C5,
B5\leftrightarrow C4, B6\leftrightarrow C3, B7\leftrightarrow C2,
B8\leftrightarrow C10\nonumber\\
&B9\leftrightarrow C9, B10\leftrightarrow C8, D1\leftrightarrow F2,
D2\leftrightarrow F1, D3\leftrightarrow F4, D4\leftrightarrow F3,
D5\leftrightarrow F5\nonumber\\
&D6\leftrightarrow F9, D7\leftrightarrow F8, D8\leftrightarrow F7,
D9\leftrightarrow F6, D10\leftrightarrow F10, E1\leftrightarrow E2,
E3\leftrightarrow E4\nonumber\\
&E5\leftrightarrow E5,  E6\leftrightarrow E9, E7\leftrightarrow E8,
E10\leftrightarrow E10
\end{eqnarray}
Note that the pattern $A1,A5,E5,E10$ transforms into itself under
the permutation. Thus among the 60 possible textures, only 32 is
independent.

\section{Numerical Analysis} We now present the numerical analysis of
the 60 neutrino mass textures of $M_{\nu}$ with one vanishing minor
and two equivalent cofactors. Our numerical calculation is performed
in the following way:

1)For each pattern, the three mixing angles $(\theta_{12},
\theta_{23}, \theta_{13})$ are allowed to vary randomly in their
$3\sigma$ range. Up to now, no bound was set on Dirac CP-violating
phase $\delta$, so we vary it randomly in the range of $[0,2\pi]$.

2)For each group of random numbers ($\theta_{12},
\theta_{23},\theta_{13}, \delta$), we calculate the corresponding
two neutrino mass ratios $\rho$ and $\sigma$. Using Eq. \eqref{rv},
the mass-squared difference ratio $R_{\nu}$ is determined. Then the
input parameters is empirically acceptable when the $R_{\nu}$ falls
inside the the $3\sigma$ range of experimental data, otherwise they
are excluded. Since we have two possible neutrino mass hierarchy, in
the analysis we further demand $\rho<\sigma<1$ corresponding to the
NH case and $\sigma<1<\rho$ corresponding to the IH case.

3)For each group of input parameter consistent from the constraint
of $R_{\nu}$ above, we randomly generate the value of $\delta m^{2}$
in its $3\sigma$ range. From Eq. \eqref{abm} ,the three neutrino
masses $(m_{1}, m_{2}, m_{3})$ are obtained. Given the three
neutrino masses, a check shall be made to figure out if the
corresponding $\Delta m^{2}$ falls into the $3\sigma$ range given in
\eqref{ms}.

4)Finally, we get the Majorana CP-violating $\alpha$ and $\beta$
though Eq.\eqref{21}, \eqref{22}. Since we have already obtained the
absolute neutrino mass $m_{1,2,3}$ and $(\alpha,\beta)$, the further
constraint from cosmology and $0\nu2\beta$ decay experiment should
be considered. A $3\sigma$ result of $\langle m
\rangle_{ee}=(0.11-0.56)$ eV is reported by the Heidelberg-Moscow
Collaboration\cite{HM}. However, this result is criticized in Ref
\cite{NND2} and shall be checked by the forthcoming experiment. In
this work, we set the upper bound on $\langle m \rangle_{ee}$ at 0.5
eV. It is believed that that the next generation $0\nu\beta\beta$
experiments, with the sensitivity of $\langle m \rangle_{ee}$ being
up to 0.01 eV, will open the window to not only the absolute
neutrino mass scale but also the Majorana-type CP violation. Besides
the $0\nu\beta\beta$ experiments, a more severe constraint was set
from the recent cosmology observation. Recently, an upper bound on
the sum of neutrino mass $\sum m_{i}<0.23$ eV is
reported\cite{Planck} by Plank Collaboration combined with the WMAP,
high-resolution CMB and BAO experiments. Once the input parameter
satisfy all the constraint given above, we give a detail discussion
on the survived pattern.

A numerical and comprehensive analysis over the sixty patterns of
$M_{\nu}$ have been carried out in our study. The main results and
the discussion are summarized as follows:

(i) Eight out of the sixty patterns, viz., A2, A3, A5, A7, A8, A9,
A10 and E5 are ruled out at $3\sigma$ confidence level, leaving
fifty-two patterns still being compatible with the current
experimental data.

(ii) Among the fifty-two surviving patterns, fifteen of them,
viz.,D3, D6, D8, E1, E2, E3, E4, E6, E7, E8, E9, E10, F4, F7 and F9
are phenomenological acceptable only for NH spectrum.

(iii) Patterns A4, A6, B8 and C10 are phenomenological acceptable
only for IH spectrum.
\begin{table}
\caption{\label{} The classification of lower bound of $\langle
m\rangle_{ee}$}
\begin{tabular}{|c|c|}
\hline \quad&
B2-C7(NH),B3-C6(NH),B8-C10(IH),B9-C9(NH),B4-C5(NH) D2-F1(NH)\\
0eV$\sim$0.01eV & ,D3-F4(NH),D4-F3(NH),D6-F9(NH),D7-F8(NH),E3-E4(NH),E6-E9(NH)\\
\quad& E7-E8(NH), E10-E10(NH),D8-F7(NH),D9-F6(NH),D10-F10(NH)\\
\hline \quad&
A1-A1(IH),A4-A6(IH),B1-C1(NH),B5-C4(NH),B6-C3(NH,IH)\\
0.01eV$\sim$0.04eV& B7-C2(NH), B9-C9(IH),B10-C8(NH,IH), D1-F2(NH), D2-F1(IH),\\
\quad& D5-F5(NH),D10-F10(IH)\\
\hline \quad & B1-C1(IH),B2-C7(IH),B3-C6(IH) B4-C5(IH),
B5-C6(IH)\\
0.04eV$\sim$0.08eV& B7-C2(IH) D1-F2(IH), D4-F3(IH),
D5-F5(IH),D7-F8(IH)\\
\quad&D9-F6(IH),E1-E2(NH)\\
\hline
\end{tabular}
\end{table}

\begin{figure}
 \includegraphics[scale = 0.70]{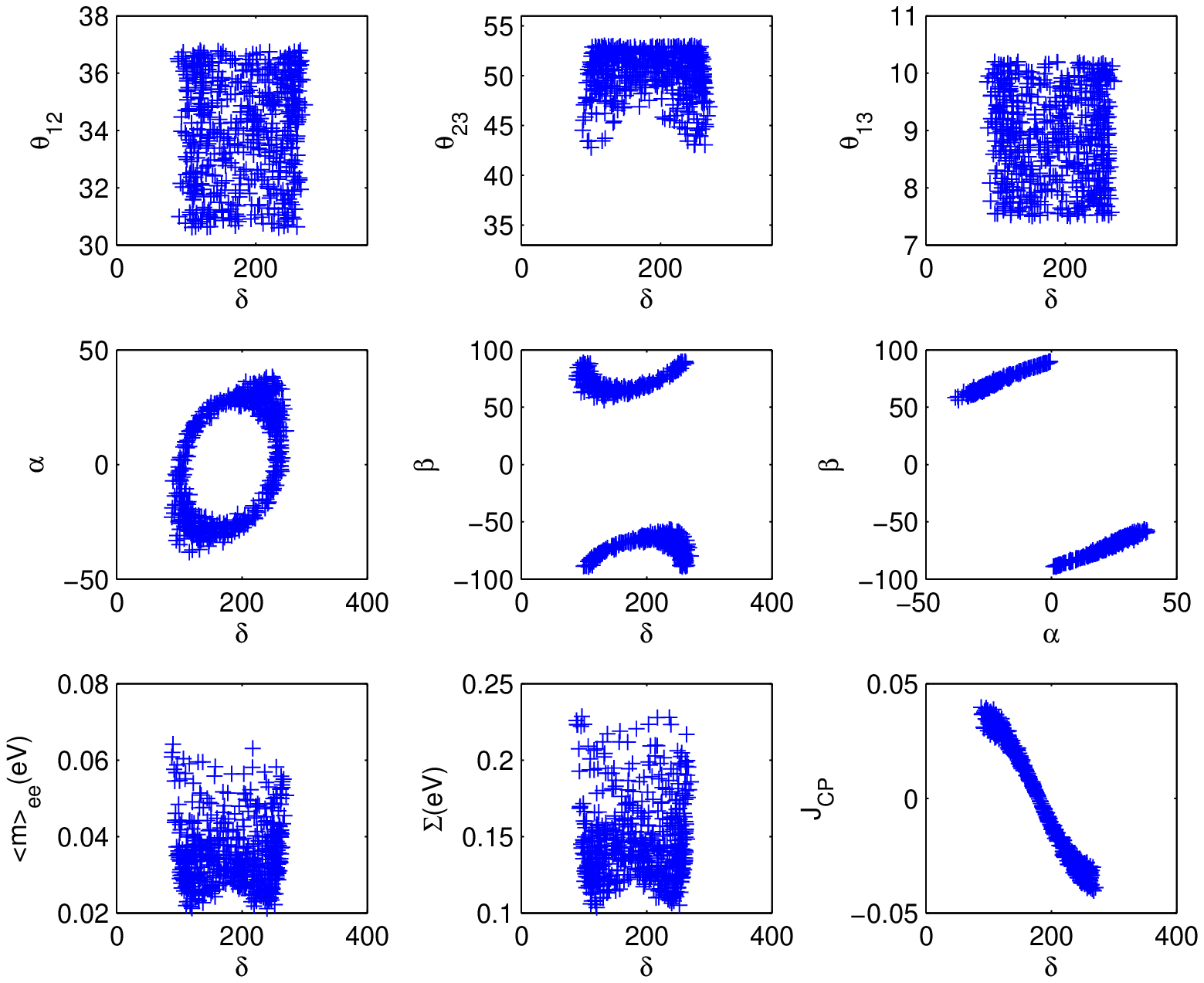}%
\caption {The plots for pattern D1 (NH). } \label{111}
 \end{figure}

\begin{figure}
 \includegraphics[scale = 0.70]{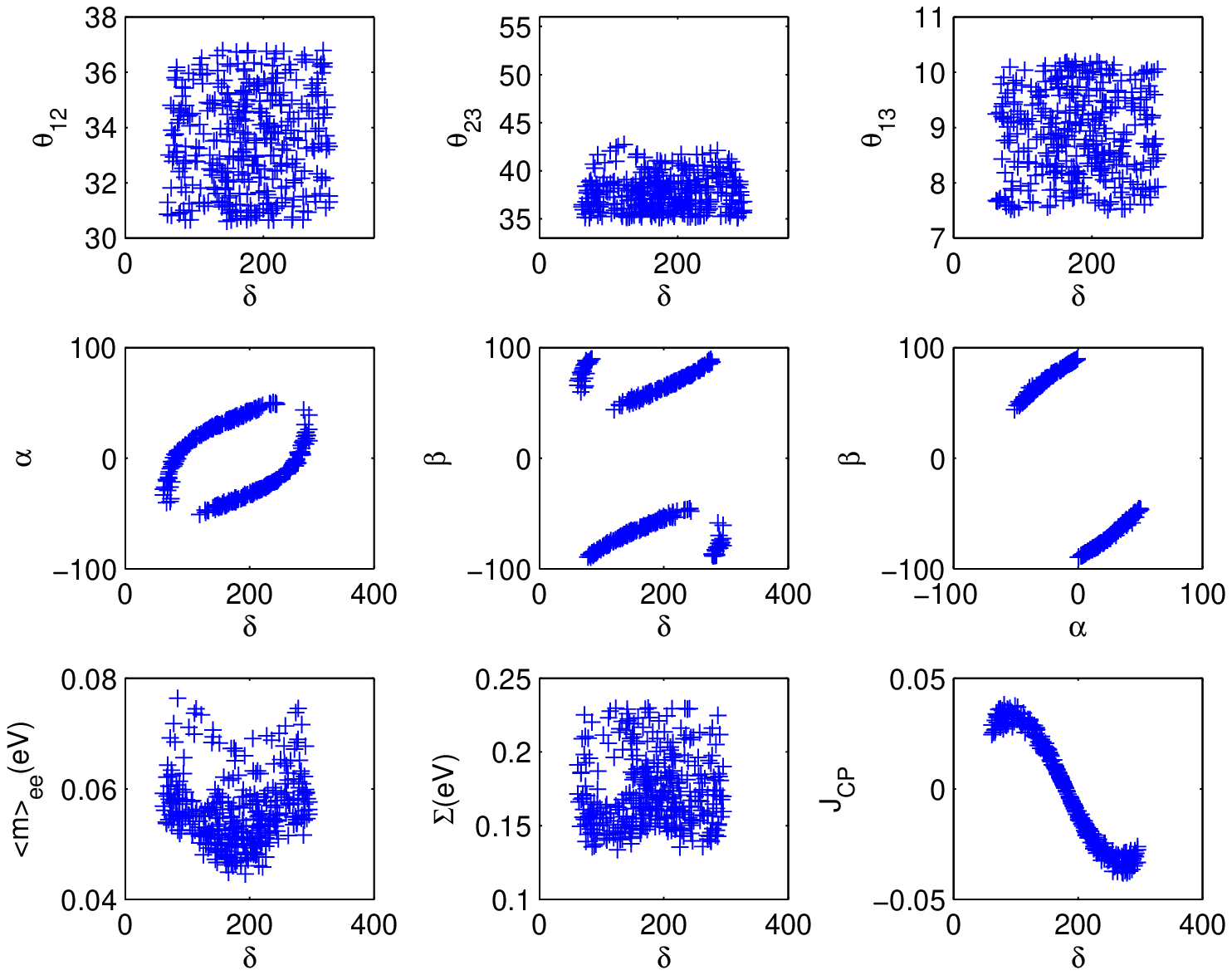}%
\caption {The plots for pattern D1 (IH). } \label{222}
 \end{figure}
(iv) The current neutrino oscillation experiments tend to give a
$\theta_{23}<45^{\circ}$ result. Using the $2\sigma$ range given in
\eqref{2sigma}, more patterns of $M_{\nu}$ will be excluded. They
are B7, C3, C8, D1, D5 for normal hierarchy and A4, A6, B3, B4, B9,
C1, C4, C7, F1, F2, F3, F5, F6, F8, F10 for inverted hierarchy.
Hence, the accurate measurement of $\theta_{23}$ and mass hierarchy
in future long-baseline neutrino oscillation experiments is
essential for our model selection.

(v) The theoretical lower bound of $\langle m\rangle_{ee}$ for
patterns are classified in Table. I. The sensitivity of next
generation $0\nu\beta\beta$ experiments is of order 0.01 eV, which
is just the same order as the $\langle m \rangle_{ee}$ of many
allowed patterns we concerned. We expect future $0\nu\beta\beta$
experiments can set a more severe bound for us to distinguish the
allowed texture. It should be emphasized that the whole range of
$\langle m\rangle_{ee}$ for textures D6-F9(NH) and D9-F6(NH) lie in
the range of [0, 0.01] which is far from the scope of forthcoming
experiment. For other textures, they all have the possibility to be
detected in the future experiment.

(vi) The upper bound $\Sigma<0.23$ is saturated for most of the
surviving patterns except for A4(IH), A6(IH), D3(NH), D6(NH),
E3(NH), E6(NH), E7(NH) and E10(NH). On the other hand for all the
allowed patterns, $\langle m \rangle_{ee}<0.080$ eV is satisfied.
Thus compared with the $\langle m \rangle_{ee}<0.5$ eV condition,
the constraint of $\Sigma$ from cosmology observation is more serve
and useful.

A full presentation of the figures for all viable textures will
render the paper unreadable. Therefore we present one of the cases,
viz., D1 pattern in Fig. \ref{111}(for NH) and Fig. \ref{222}(for
IH) as concrete demonstrations.

In both Fig. \ref{111} and Fig. \ref{222}, we have plotted the
allowed ranges of the mixing angle $\theta_{12}$, $\theta_{23}$ and
$\theta_{13}$ as the function of Dirac-CP violating phase $\delta$
for the D1 pattern in the first row. We have seen that the different
type of mass hierarchy shows different feature. For the IH spectrum,
although accepted in $3\sigma$ range, the predicted $\theta_{23}$ is
disallowed by $2\sigma$ range. For the IH spectrum, however, the
predicted $\theta_{23}$ is accepted by both $3\sigma$ and $2\sigma$
ranges. We present the ranges of Majorana-CP violating phase
$(\alpha, \beta)$ versus $\delta$ and the $\alpha- \beta$
correlation plots in the second row, where an approximately linear
correlation between $\alpha$ and $\beta$ is existed for both NH and
IH spectrum. In the third row, we have showed the $0\nu \beta\beta$
decay $\langle m \rangle_{ee}$, the sum of neutrino mass $\Sigma$
and the Jarlskog invariant with respect of $\delta$. One can observe
from the figure that the lower bound of $\langle m \rangle_{ee}$ of
D1(NH) and D1(IH) textures are approximately located at 0.020 eV and
0.045 eV, which is in accordance with the results given in Table. I
and reachable in the future $0\nu\beta\beta$ decay experiments.
Furthermore, the sum of neutrino masses $\Sigma\geq0.1$ eV are
satisfied for both NH and IH hierarchy, leading to quasidegenerate
spectrums($m_{i}\approx m_{j}\gg m_{i}-m_{j}$). On the other hand,
the Jarlskog invariant is not strongly constrained.  The Jarlskog
invariant can reach its maximal value, $\mid J_{CP}\mid\sim4\%$ if
the Dirac-CP phase $\delta$ is close to $\pi /2$ and $3\pi/2$ for
normal as well as inverted hierarchy.

\section{Symmetry Realization}
It is generally believed that the observed neutrino mixing pattern
suggests some underlying discrete flavor symmetries (for a review,
see\cite{fla}). It has been proved that the hybrid $M_{\nu}$
textures can be achieved by imposing the $S_{3}\otimes Z_{3}$ flavor
symmetry in type-II seesaw model\cite{hybrid2}. Thus, it is natural
to ask if the hybrid $M_{\nu}^{-1}$ textures we studied can arise
from this flavor symmetry. As an illustration, We propose a flavor
model for D4 pattern. Consider the three left-handed neutrino fields
$\nu_{iL}$, three right-handed neutrino fields $\nu_{Ri}$ where
$i=e,\mu, \tau$ in the following representation of $S_{3}\otimes
Z_{3}$ group
\begin{equation}
l_{\mu L}\sim (1,\omega^{2}), \quad \left(\begin{array}{c}
l_{e L}\\
l_{\tau L}
\end{array}\right)\sim (2,\omega), \quad \nu_{\mu
R}\sim(1,\omega),\quad \left(\begin{array}{c}
\nu_{e R}\\
\nu_{\tau R}
\end{array}\right)\sim (2,\omega^{2})
\end{equation}
and for scalar fields
\begin{equation}
\left(\begin{array}{c}
\chi_{1}\\
\chi_{2}
\end{array}\right)\sim (2,1), \quad \left(\begin{array}{c}
\chi_{3}\\
\chi_{4}
\end{array}\right)\sim (2,\omega^{2}),\quad \chi_{5}\sim(1,\omega^{2}),\quad
\Phi\sim(1,1)
\end{equation}
where $\omega=e^{i2\pi\setminus3}$Then the Lagrangian corresponding
to the Dirac neutrino mass term and Majorana neutrino mass term are
given by
\begin{eqnarray}
L_{D}=Y_{1}(\overline{l}_{\mu L}\nu_{\mu
R})\Phi+Y_{2}(\overline{l}_{e L}\nu_{e
R}+\overline{l}_{\tau L}\nu_{\tau R})\Phi+ h.c\\
L_{M}=\frac{1}{2}y_{1}\nu_{\mu R}^{T}C^{-1}(\chi_{1}\nu_{e
R}+\chi_{2}\nu_{\tau R})-\frac{1}{2}y_{2}(\nu_{e R}^{T}C^{-1}\nu_{e
R}+\nu_{\tau R}^{T}C^{-1}\nu_{\tau R})\chi_{5}\nonumber\\
-\frac{1}{2}y_{3}[(\nu_{e R}^{T}C^{-1}\nu_{\tau R}+\nu_{\tau
R}^{T}C^{-1}\nu_{e R})\chi_{3}+(\nu_{e R}^{T}C^{-1}\nu_{e
R}-\nu_{\tau R}^{T}C^{-1}\nu_{\tau R})\chi_{4}]+h.c
\end{eqnarray}
where the direct product $2\otimes2=1+1^{'}+2$ for $S_{3}$ group is
used\cite{fla2}. When the Higgs fields acquire the vacuum
expectations values (VEVs) $\langle \Phi \rangle=v$, $\langle
\chi_{i} \rangle=v_{i}$, we obtain the Dirac mass matrix and the
Majorana mass matrix as
\begin{equation}
M_{D}=\left(\begin{array}{ccc}
Y_{2}v&0&0\\
0&Y_{1}v&0\\
0&0&Y_{2}v
\end{array}\right), \quad M_{R}=\left(\begin{array}{ccc}
y_{2}v_{5}+y_{3}v_{4}&y_{1}v_{1}&2y_{3}v_{3}\\
y_{1}v_{1}&0&y_{1}v_{2}\\
2y_{3}v_{3}&y_{1}v_{2}&y_{2}v_{5}-y_{3}v_{4}
\end{array}\right)
\end{equation}
Once the $v_{4}=0$ after the scalar potential is minimized, one can
check that the neutrino mass matrix
$M_{\nu}=M_{D}M_{R}^{-1}M_{D}^{T}$ satisfies the condition
$C_{22}=0$ and $C_{11}=C_{33}$ of D4 pattern.

On the other hand, the charged lepton mass matrix $M_{l}$ should be
diagonal. The right-handed charged lepton are assigned as
\begin{equation}
\mu_{R}\sim (1,\omega^{2}),\quad e_{R}\sim (1^{\prime},\omega),\quad
\tau_{R}\sim (1,\omega)
\end{equation}
To generate the charged lepton mass, we still need a standard Higgs
field which is a singlet scalar under the flavor symmetry and scalar
doublet fields $\Phi_{1}$ and $\Phi_{2}$
\begin{equation}
H\sim (1,1),\quad \left(\begin{array}{c}
\Phi_{1}\\
\Phi_{2}
\end{array}\right)\sim (2,1),
\end{equation}
The Lagrangian of charged lepton sector is given by
\begin{equation}
L_{l}=Y_{\mu}\overline{l}_{\mu
L}\mu_{R}H+Y_{e}(\overline{l}_{eL}\Phi_{2}-\overline{l}_{\tau
L}\Phi_{1})e_{R}+Y_{\tau}(\overline{l}_{e
L}\Phi_{1}+\overline{l}_{\tau L}\Phi_{2})\tau_{R}+h.c
\end{equation}
After the vacuum expectation $\langle H\rangle=v_{6}$, $\langle
\Phi_{2}\rangle=v_{7}$ and $\langle \Phi_{1}\rangle=0$ are taken, we
obtain the charged lepton mass matrix $M_{l}=Diag\{m_{e}, m_{\mu},
m_{\tau}\}$ with $m_{e}=Y_{e}v_{7}$, $m_{\mu}=Y_{\mu}v_{6}$,
$m_{\tau}=Y_{\tau}v_{7}$.

Hence we construct the a hybrid $M_{\nu}^{-1}$ neutrino mass texture
via $S_{3}\otimes Z_{3}$ flavor symmetry. However, it should be
stressed that there still remain two things unsolved: $1)$, it is
not trivial to guarantee the Higgs fields are broken to a specific
direction. i.e. the vacuum alignment problem; $2)$. As similar as
the hybrid $M_{\nu}$ ones, not all the textures can be realized by
$S_{3}\otimes Z_{3}$ flavor symmetry. However, tt is noted that in
another work\cite{new}, the $Z_{2}\otimes Z_{4}$ symmetry is
produced to generate the texture
\begin{equation}
\left(\begin{array}{ccc}
0&\bigtriangleup&\bigtriangleup\\
\bigtriangleup&\times&\times\\
\bigtriangleup&\times&\times
\end{array}\right)
\end{equation}
which is just the A1 pattern.

In this sense, the symmetry realization for all neutrino mass
patterns in a systematic and self-consistent way deserves further
research.

\section{Conclusion and Discussion}
In this work, we have presented a numerical and comprehensive study
of the neutrino mass textures with one vanishing minor and two equal
cofactors. Among the sixty texture, only eight of them are ruled out
for NH as well as IH spectrum by the current experiments data at
$3\sigma$ level. The neutrinoless double decay experiments and the
neutrino sum mass from cosmology observation are also discussed. We
expect that further experiment will provide us more accurate
determination on the mixing angle, CP-violating phase, mass
hierarchy and neutrino absolute mass, and will finally help us
select the appropriate structure of mass texture. In this processes,
the determination of $\theta_{23}$ plays an essential role on the
model selection. A flavor realization based on $S_{3}\otimes Z_{3}$
symmetry is also discussed. The symmetry realization of all the
textures in a systematic and self-consistent way deserves further
research. We except that a cooperation between theoretical study
from the flavor symmetry point view and a phenomenology study will
help us reveal the structure of neutrino mass texture.

\begin{acknowledgments}
The author would like to thank S. Dev, Zheng-Mao Sheng, Jia-Hui
Huang, Ji-Yuan, Liu, and Shun Zhou for the useful discussion during
this work.
\end{acknowledgments}


\begin{thebibliography}{21}
\bibitem[1]{neu1}
Q.R. Ahmad \textsl{et al.}(SNO Collaboration), Phys. Rev. Lett {\bf
89}, 011301(2002); K. Eguchi \textsl{et al.} (KamLAND
Collaboration), Phys. Rev. Lett {\bf 90}, 021802(2003); M.H. Ahn
\textsl{et al.} (K2K Collaboration), Phys. Rev. Lett {\bf 90},
041801(2003).

\bibitem[2]{neu2}
F.P. An \textsl{et al.} (DAYA-BAY Xollaboration), Phys. Rev. Lett.
{\bf 108}, 171803(2012).

\bibitem[3]{neu3}
J.K. Ahn \textsl{et al.} (RENO Collaboration), Phys. Rev. {\bf
D108}, 191802(2012).

\bibitem[4]{WMAP}
G. Hinshaw \textsl{et al.} (WMAP Collaboration), arXiv: 1212.5226.

\bibitem[5]{Planck}
P.A.R. Ade \textsl{et al.} (Planck Collaboration), arXiv: 1303.5076.

\bibitem[6]{NDD}
S.M. Bilenky and C. Giunti, Mod. Phys, Lett. {\bf A16},
1230015(2012).

\bibitem[7]{zero}
P.H. Frampton, S. L. Glashow, and D. Marfatia, Phys. Lett. {\bf
B536}, 79(2002); Z,-z. Xing, Phys. Lett. {\bf B530}, 159(2002); M.
Randhawa, G. Ahuja, and M. Gupta, Phys. Lett. {\bf B643}, 175(2006);
A. Merle, and W. Rodejohann, Phys. Rev. {\bf D73}, 073012(2006); S.
Dev, S. Kumar, S. Verma, and S. Gupta, Phys. Rev. {\bf D76},
013002(2007); S. Dev, S. Kumar, S. Verma, and S. Gupta, Nucl. Phys.
{\bf B784}, 103(2007); G. Ahuja, S. Kumar, M. Randhawa, M. Gupta,
and S. Dev, Phys. Rev. {\bf D76}, 013006(2007); S. Dev, S. Kumar,
Mod. Phys, Lett. {\bf A22}, 1401(2007); S. Kumar, Phys. Rev. {\bf
D84}, 077301(2011); P.O. Ludl, S. Morisi, and E. Peinado, Nucl.
Phys. {\bf B857}, 411(2012); W. Grimus. and P.O. Ludl,
arXiv:1208.4515; D. Meloni, and G. Blankenburg, Nucl. Phys. {\bf
B867}, 749(2013); H. Fritzsch, Z.-z. Xing, and S. Zhou, J. High
Energy Phys. 09 (2011)083.

\bibitem[8]{hybrid}
S. Kaneko, H. Sawanaka, and M. Tanimoto, J. High Energy Phys. 08
(2005)073; S. Dev, S. Verma, and S. Gupta, Phys. Lett. {\bf B687},
53(2010); S. Dev, S. Gupta, and R.R. Gautam, Phys. Rev. {\bf D82},
073015(2010); S. Goswami, S. Khan, and A. Watanable, Phys. Lett.
{\bf B687}, 53(2010), W. Grimus, and P. O. Ludl, arXiv: 1208.4515.

\bibitem[9]{hybrid2}
J.-Y. Liu and S. Zhou, Phys. Rev. {\bf D87}, 093010(2013).


\bibitem[10]{sum}
X.-G. He and A. Zee, Phys. Rev. {\bf D68}, 037302(2003).

\bibitem[11]{det}
G.C. Branco, R. Gonzalez Felipe, F.R. Joaquim, and T. Yanagida,
Phys. Lett. {\bf B562}, 265(2003); B.C. Chauhan, J. Pulido, and M.
Picariello, Phys. Rev. {\bf D73}, 053003(2006).

\bibitem[12]{minor}
L. Lavoura, Phys. Lett. {\bf B609}, 317(2005); E.I. Lashin and N.
Chamoun, Phys. Rev. {\bf D78}, 073002(2008); E.I. Lashin and N.
Chamoun, Phys. Rev. {\bf D80}, 093004(2009);  S. Dev, S. Verma, S.
Gupta, and R.R. Gautam, Phys. Rev. {\bf D81}, 053010(2010); S. Dev,
S. Gupta, and R.R. Gautam, Mod. Phys, Lett. {\bf A26}, 501(2011); S.
Dev, S. Gupta, R.R. Gautam, and L. Singh, Phys. Lett. {\bf B706},
168(2011); T. Araki, J. Heeck, and J. Kubo, J. High Energy Phys. 07
(2012)083; S. Verma, Nucl. Phys. {\bf B854}, 340(2012).

\bibitem[13]{tra}
H.A. Alhendi, E.I. Lashin, and A.A. Mudlej, Phys. Rev. {\bf D77},
013009(2008).

\bibitem[14]{co}
S. Dev, R.R. Gautam, and L. Singh, Phys. Rev. {\bf D87},
073011(2013).

\bibitem[15]{seesaw}
H. Fritzsch, M. Gell-Mann, and P. Minkowski, Phys. Lett. {\bf B59},
256(1975); P. Minkowski, Phys. Lett. {\bf B67}, 421(1977); T.
Yanagida, in \textsl{Proceedings of Workshop on Unified Theory and
the Baryon Number of the Universe}, edited by O. Sawada and A.
Sugamoto(KEK, Tsukuba, 1979), p. 95; M. Gell-Mann, P. Ramond, and
Slansky, in \textsl{Supergravity}, edited by P. van. Nieuwenhuizen
and D.Z. Freeman (North-Holland, Amsterdam,1979), p. 315; R.N.
Mohapatra and G. Senjanovic, Phys, Rev. Lett. {\bf 44}, 912(1980);
J. Schechter and J. W. F. Valle,  Phys. Rev. {\bf D22}, 2227(1980);
J. Schechter and J. W. F. Valle,  Phys. Rev. {\bf D25}, 774(1982).


\bibitem[16]{Ma}
E. Ma, Phys. Rev. {\bf D71}, 111301(2005).

\bibitem[17]{Fog}
G.L. Fogli, E. Lisi, A. Marrone, and A. Palazzo, Prog. Part. Nucl.
Phys. {\bf 57}, 742(2006).

\bibitem[18]{PMNS}
B. Pontecorvo, Zh. Eksp. Teor. Fiz. {\bf 33}, 549(1957); Z. Maki, M.
Nakagawa, and N. Sakata, Prog. Theor. Phys. {\bf 28}, 870(1962).

\bibitem[19]{Jas}
C. Jarlskog,  Phys, Rev. Lett. {\bf 55}, 1039(1985).

\bibitem[20]{data}
G.L. Fogli, E. Lisi, A. Marrone, D. Montanino, A. Palazzo, and A.M.
Rotunno, Phys. Rev. {\bf D86}, 013012(2012); D. V. Forero, M.
Tortola, and J. W. F. Valle, Phys. Rev. {\bf D86}, 073012(2012); M.
C. Gonzalez-Garcia, Michele Maltoni, Jordi Salvado, and Thomas
Schwetz, J. High Energy Phys. 12 (2012)123.

\bibitem[21]{mut}
T. Fukuyama and H. Nishiura, arXiv: 9702253; R. N. Mohapatra and S.
Nussinov, Phys. Rev. D60, 013002, (1999); E. Ma and M. Raidal, Phys.
Rev. Lett. 87, 011802 (2001); C. S. Lam, Phys. Lett. B507, 214
(2001); K. R. S. Balaji, W. Grimus and T. Schwetz, Phys. Lett. B508,
301 (2001); W. Grimus and L. Lavoura, Acta Phys. Pol. B32, 3719
(2001).
\bibitem[22]{HM}
H.V. Klapdor-Kleingrothaus, A. Dietz, H.L. Harney, and I.V.
Krivosheina, Mod. Phys. Lett. {\bf A16}, 2409(2001).

\bibitem[23]{NND2}
C.E. Aalseth \textsl{et al}. Mod. Phys. Lett. {\bf A17}, 1475(2002);
F. Feruglio, A. Strumia, and F. Vissani, Nucl. Phys. {\bf B637},
345(2002).

\bibitem[24]{fla}
G. Altarelli and F. Feruglio, Rev. Mod. Phys. {\bf 82}, 2701(2010);
S. K. King and C, Luhn, Rept.Prog.Phys. 76 (2013) 056201.

\bibitem[25]{fla2}
H. Ishimori, T. Kobayashi, H. Ohki, Y. Shimizu, H. Okada, and M.
Tanimoto, Prog. Theor. Phys. Suppl. {\bf 183} 1(2010).

\bibitem[26]{new}
S. Dev, R. R. Gautam, L. Singh, arXiv: 1306.4281.
\end{thebibliography}
\end{document}